\documentclass[twocolumn,english]{article}
\usepackage{bookman}
\usepackage[T1]{fontenc}
\usepackage[latin1]{inputenc}
\usepackage{color}
\usepackage{graphicx}
\usepackage{amssymb}

\makeatletter

\providecommand{\tabularnewline}{\\}

\usepackage{babel}
\makeatother
\begin{document}
\begin{minipage}[c]{1.0\textwidth}%
\begin{center}\textbf{\Large How Nuclear Diffuseness Affects RHIC
Data}\end{center}{\Large \par}

\noindent \begin{center}\textbf{\large Klaus WERNER }\end{center}{\large \par}

\noindent \begin{center}\textit{SUBATECH, Université de Nantes --
IN2P3/CNRS-- EMN,  Nantes, France}\emph{}\\
\emph{werner@subatech.in2p3.fr}\end{center}

Abstract: The fact that nuclei have diffuse surfaces (rather than
being simple spheres) has dramatic consequences on the interpretation
of the RHIC heavy-ion data. The effect is quite small (but not negligible)
for central collisions, but gets increasingly important with decreasing
centrality. One may actually divide the collision zone into a central
part ({}``core''), with expected high energy densities, and a peripheral
part ({}``corona''), with smaller energy densities, more like in
pp or pA collisions. We will discuss that many complicated {}``features''
observed at RHIC become almost trivial after subtracting the corona
background. We are focussing on AuAu collisions at 200 GeV.\end{minipage}%

Everybody having worked on pA scattering knows about the importance
of the nuclear diffuseness. Even though the proton suffers on the
average up to 6 collisions when it traverses a big nucleus, the most
likely situation is still just one interaction, due to the surface
diffuseness.

In nuclear collisions, the surface effect is as well present, and
also very important. The peripheral nucleons of either nucleus essentially
perform independent pp or pA-like interactions, with a very different
particle production compared to the high density central part. For
certain observables, this {}``background'' contribution completely
spoils the {}``signal'', and to properly interpret RHIC data, we
need to subtract this background.

In order to get quantitative results, we need a simulation tool, and
here we take EPOS \cite{epos}, which has proven to work very well
for pp and dAu collisions at RHIC. \textbf{It is very important to
understand that the main results of this paper do not depend on whether
or not the model treats the high density part 100\% correctly. The
crucial point is that the model describes pp and pAu to a high precision,
so we can safely subtract the peripheral low density part.}

\textbf{\vspace*{6.5cm}}

EPOS is a parton model, so in case of a AuAu collision there are many
binary interactions, creating partons, which then hadronize employing
a phenomenological procedure call string fragmentation. Here, we modify
the procedure: we have a look at the situation at an early proper
time $\tau_{0}$, long before the hadrons are formed: we distinguish
between string segments in dense areas (more than $\rho_{0}$ segments
per unit proper volume), from those in low density areas. In the following,
we will use $\tau_{0}=1$ and $\rho_{0}=1\,\mathrm{f}\mathrm{m}^{-3}$.
We refer to high density areas as core, and to low density areas as
corona.

\begin{figure}
~

\includegraphics[%
  scale=0.35,
  angle=270]{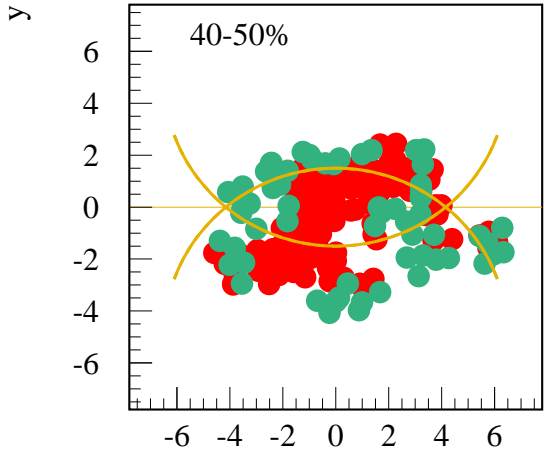}

\includegraphics[%
  scale=0.35,
  angle=270]{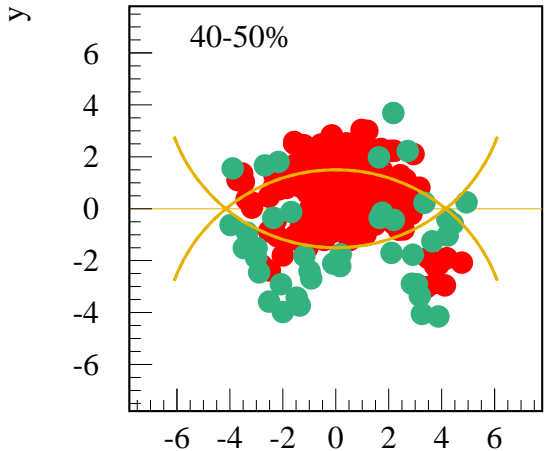}

\caption{Two Monte Carlo realizations of semi-peripheral (40-50\%) AuAu collisions
at 200 GeV (cms). We show in red string segments in high density areas
(core), and in green the string segments in low density environments
(corona). The circles are put in just to guide the eye: they represent
the two nuclei in hard sphere approximation.We consider a projection
of segments within $z=\pm0.8\,\mathrm{fm}$ to the transverse plane
(x,y).\label{cap:geom1}}
\end{figure}

In figure. \ref{cap:geom1}, we show two examples (randomly chosen)
of semi-peripheral (40-50\%) AuAu collisions at 200 GeV (cms), simulated
with EPOS.%
We observe large fluctuations event by event, simply reflecting the
randomness of the positions of binary nucleon-nucleon collisions.
There is an important contribution from the low density area, contributing
roughly 20\% to the final particle production. But much more importantly,
as discussed later, the importance of this contribution depends strongly
on particle type and transverse momentum. For central collisions,
the low density contribution is obviously less important, for more
peripheral collisions this contribution will even dominate.

How do these low density contributions interact with the expanding
core? Well, even a system of noninteracting particles expands, with
the velocity of light (reflecting the outward moving particles ).
Inward moving particles may be absorbed by the core, on the other
hand the core edges start to hadronize at the same time, with a good
chance that early hadronization and absorption compensate each other.
So we ignore any interaction for the moment. But even if part of the
low density contribution will be absorbed, there will be a sizable
effect.

In order to make a quantitative statement, we adopt the following
strategy: the low density part will be treated using the usual EPOS
particle production which has proven to be very successful in pp and
dAu scattering (the peripheral interactions are essentially pp or
pA scatterings). For the high density part, we simply try to parameterize
particle production, in the most simple way possible (It is not at
all our aim to provide a microscopic description of this part). Suppose
we find such a simple parameterization of the core contribution, such
that the total contribution reproduces all the relevant low and intermediate
$p_{t}$ data, then our core parameterization represents in fact the
data after {}``background subtraction'', and that is what we are
really interested in!

In practice, we first divide the EPOS string segments into core and
corona contribution, as discussed earlier (apart of the density, there
is another condition: only segments with transverse momenta less than
4 GeV contribute to the core, the others escape freely, no jet quenching).
We then consider the core contributions more closely, in longitudinal
slices, characterized by some range in $\eta=0.5\ln(t+z)/(t-z)$.
Since string segments show a Bjorken-fluid-like behavior, the particles
in a segments around $\eta$ move with rapidities close to $\eta$.
Connected core regions in a given segment are considered to be clusters,
whose energy and flavor content are complete determined by the corresponding
string segments. Clusters are then considered to be collectively expanding:
Bjorken-like in longitudinal direction, with in addition some transverse
expansion. We assume that the clusters hadronize at some given energy
density $\varepsilon_{\mathrm{hadr}}$, having acquired at that moment
a collective radial flow, with a linear radial rapidity profile from
inside to outside, characterized by the maximal radial rapidity $y_{\mathrm{rad}}$.
In addition, we impose an azimuthal asymmetry, by multiplying the
$x$ and $y$ component of the flow four-velocity with $1+\epsilon\, f_{\mathrm{ecc}}$
and $1-\epsilon\, f_{\mathrm{ecc}}$, where $\epsilon$ is the the
initial spacial eccentricity, $\epsilon=\left\langle (y^{2}-x^{2})/y^{2}+x^{2})\right\rangle $,
and $f_{\mathrm{ecc}}$ a parameter. By imposing radial flow, we have
to rescale the cluster mass as\[
M\to M\,\times\,0.5\, y_{\mathrm{rad}}^{2}/(y_{\mathrm{rad}}\sinh y_{\mathrm{rad}}-\cosh y_{\mathrm{rad}}+1),\]
in order to conserve energy. Hadronization then occurs according to
covariant phase space, which means that the probability $dP$ of a
given final state of n hadrons is given as\textcolor{black}{\[
\hspace{-0.8cm}\prod_{\mathrm{species}\,\alpha}\!\!\!{\frac{1}{n_{\alpha}!}}\prod_{i=1}^{n}\,\frac{d^{3}p_{i}}{(2\pi\hbar)^{3}2E}g_{i}\, s_{i}\, W\,\delta(E-\Sigma E_{i})\,\delta(\Sigma\vec{p}_{i})\,\delta_{f,\Sigma f_{i}},\]
with $p_{i}=(E_{i},\vec{p_{i}})$ being the four-momentum of the i-th
hadron, $g_{i}$ its degeneracy, and $f_{i}$ its quark flavor content
($u-\bar{u},$$d-\bar{d}$...). There is a factor $s_{i}=\gamma_{s}\,^{\pm1}$
for each strange particle (sign plus for a baryon, sign minus for
a meson), with $\gamma_{s}$ being a parameter. The number $n_{\alpha}$
counts the number of hadrons of species $\alpha$. $E$ is the total
energy of the cluster in its cms, $W$ is the cluster proper volume.
The whole procedure perfectly conserves energy, momentum, and flavors
(microcanonical procedure).}

So the core definition and its hadronization are parameterized in
terms of 6 global parameters:

\begin{center}\begin{tabular}{|c|c|c|}
\hline 
$\tau_{0}$&
 1 fm&
core formation time\tabularnewline
\hline 
$\rho_{0}$&
1 $\mathrm{f}\mathrm{m}^{-3}$ &
core formation density\tabularnewline
\hline 
$\varepsilon_{\mathrm{hadr}}$&
0.22 $\frac{GeV}{\mathrm{f}\mathrm{m}^{3}}$&
hadr. energy density\tabularnewline
\hline 
$y_{\mathrm{rad}}$&
0.83&
max. radial flow rapidity\tabularnewline
\hline 
$f_{\mathrm{ecc}}$&
0.5&
eccentricity coefficient\tabularnewline
\hline 
$\gamma_{s}$&
1.3&
hadronization factor\tabularnewline
\hline
\end{tabular}\end{center}

\noindent The final results are insensitive to variations of $\tau_{0}$,
even changes as big as a factor of 2 do not affect the results at
all. This is a nice feature, indicating that the very details of the
initial state do not matter so much. We call these parameters {}``global'',
since they account for all observables at all possible different centralities
at RHIC. In the following, we are going to discuss results, all obtained
with the above set of parameters.

\begin{figure}
\begin{center}\includegraphics[%
  scale=0.3,
  angle=270]{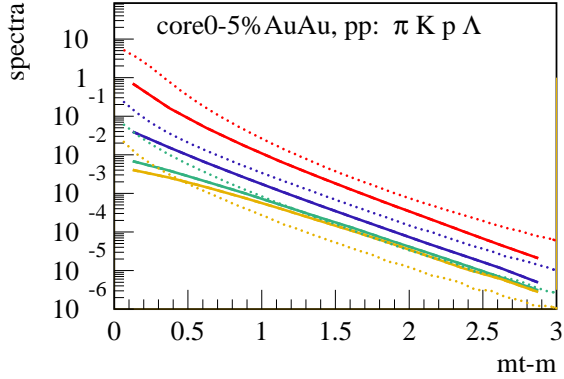}\end{center}
\vspace{-1.5cm}

\caption{Invariant yields $1/2\pi m_{t}\, dn/dydm_{t}$ of pions (red), kaons
(blue), protons (green), and lambdas (yellow), for the core contribution
corresponding to a central (0-5\%) AuAu collision (full lines) and
proton-proton scattering (dashed).\label{cap:core-pp}}
\end{figure}

All the discussion of RHIC data will be based on the interplay between
core and corona contributions. To get some feeling, we first compare
in fig. \ref{cap:core-pp} the core contribution corresponding to
a central (0-5\%) AuAu collision (which means purely statistical hadronization,
with flow) with pp scattering (which is qualitatively very similar
to the corona contribution). We plot $m_{t}$ spectra of pions, kaons,
protons, and lambdas, the nuclear spectra are divided by the number
of binary collisions (according to Glauber). We observe several remarkable
features: the shapes of the different pp spectra are not so different
among each each other, there is much more species dependence in the
core spectra, since the heavier particles acquire large transverse
momenta due to the flow effect. The second main observation concerns
the yields, in particular at intermediate values of $m_{t}-m$: the
yields for the different pp contributions are much wider spread than
the core contributions; in particular, pion production is suppressed
in the core hadronization compared to pp, whereas lambda production
is favored. All this is quite trivial, but several {}``mysteries''
discussed in the literature (and to be discussed later in this paper)
are just due to this.

\begin{figure}
\begin{center}\includegraphics[%
  scale=0.3,
  angle=270]{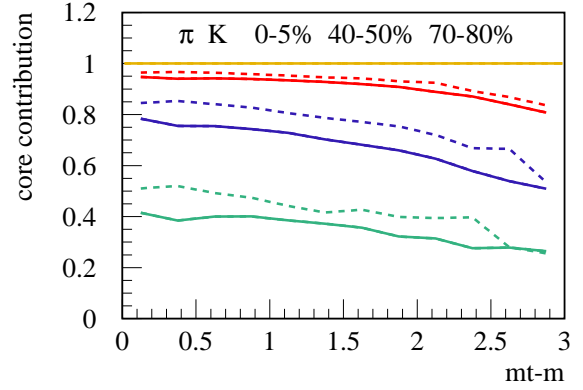}\end{center}
\vspace{-2cm}

\begin{center}\includegraphics[%
  scale=0.3,
  angle=270]{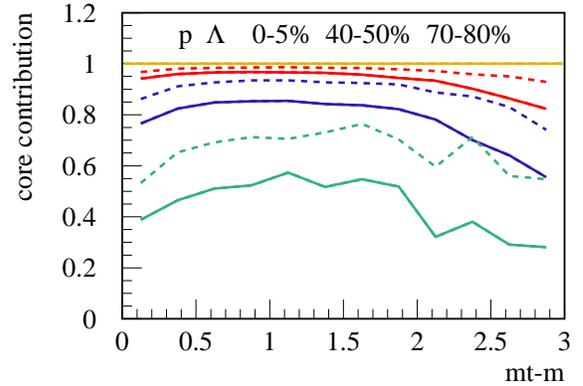}\end{center}
\vspace{-1.5cm}

\caption{The relative contribution of the core (core/(core+corona)) as a function
of the transverse mass for different centralities (0-5\%: red, 40-50\%:
blue, 70-80\%: green). Upper figure: pions (full) and kaons (dashed).
Lower figure: protons (full) and lambdas (dashed).\label{cap:core-corona}}
\end{figure}

Let us now compare core and corona contributions for different centralities
in AuAu collisions at 200 GeV. In fig. \ref{cap:core-corona}, we
plot the relative contribution of the core (relative to the complete
spectrum, core + corona) as a function of $m_{t}-m$, for different
particle species.%
For central collisions, the core contribution dominates largely (around
90\%), whereas for semi-central collisions (40-50\%) and even more
for peripheral collisions the core contribution decreases, giving
more and more space for the corona part. Apart of these general statements,
the precise $m_{t}$ dependence of the relative weight of core versus
corona depends on the particle type, and can be easily understood
by inspecting figure \ref{cap:core-pp}, since the corona contribution
is up to a factor very close to pp.

\begin{figure}
\begin{center}\includegraphics[%
  scale=0.3,
  angle=270]{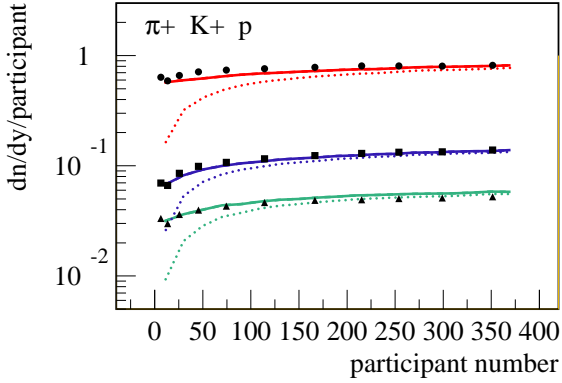}\end{center}
\vspace{-1.5cm}

\caption{Rapidity density dn/dy per participant as a function of the number
of participants, for $\pi^{+}$ (red), $K^{+}$ (blue), and $p$ (green).
We show data (points) \cite{phenix} together with the full calculation
(core + corona, full line) and just the core part (dashed).\label{cap:centrality}}
\end{figure}

We are now ready to investigate RHIC data. In fig. \ref{cap:centrality},
we plot the centrality dependence of the particle yield per participant
(per unit of rapidity), for $\pi^{+}$, $K^{+}$, and $p$, the data
together with the full calculation, but also indicating the core contribution.
The complete calculation follows quite closely the data. Whereas central
collisions are clearly core dominated, the core contributes less and
less with decreasing centrality. Similar results are obtained for
$\pi^{-}$, $K^{-}$, and $\bar{p}$, and also lambdas and xis. 

\begin{figure}
\begin{center}\includegraphics[%
  scale=0.3,
  angle=270]{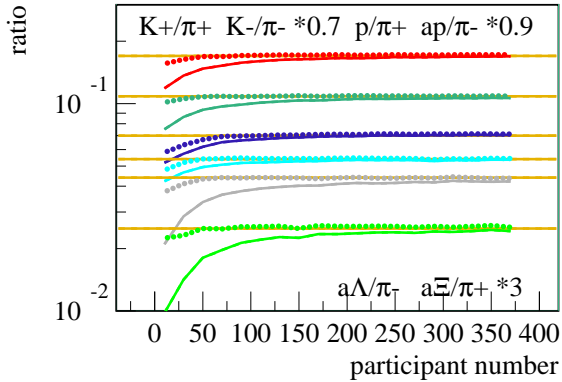}\end{center}
\vspace{-1.5cm}

\caption{Particle ratios as a function of centrality: $K^{+}/\pi^{+}$(red),
$K^{-}/\pi^{-}$(green), $p/\pi^{+}$ (blue), $\bar{p}/\pi^{-}$(cyan),
$\Lambda/\pi^{-}$(gray), $\Xi^{-}/\pi^{-}$. Complete calculation
(full) and just ratio of the core contributions (dotted). \label{cap:ratios}}
\end{figure}

Next we consider particle ratios, as a function of centrality. In
fig. \ref{cap:ratios}, we show the ratios of different particles,
with respect to pions. Whereas the complete contribution (as the data)
show a strong centrality dependence, the rations are practically flat
for the core contributions, apart of some decrease for very small
participant numbers.%

\textbf{So our first important conclusion: after subtracting the {}``corona
background'', the interesting part, the core contribution, shows
an extremely simple behavior: there is no centrality dependence, the
systems are simply changing in size (and the participant number is
certainly not a good measure of the volume of the core part, this
is why the overall multiplicities per participant decrease with decreasing
centrality).}

\begin{figure}
\begin{center}\includegraphics[%
  scale=0.3,
  angle=270]{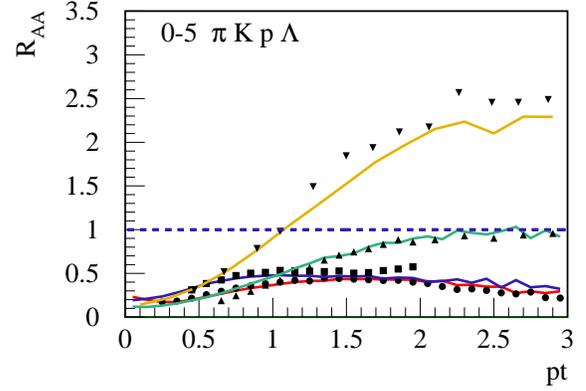}\end{center}
\vspace{-1.5cm}

\caption{Nuclear modification factors in central AuAu collisions at 200 GeV.
Lines are full calculations, symbols represent data \cite{phenix,star-lda}.
We show results for pions (red; circles), kaons (blue; squares), protons
(green; triangles), and lambdas (yellow; inverted triangles). \label{cap:nmf}}
\end{figure}

Lets us come to $p_{t}$ or $m_{t}$ spectra. We checked all available
low and intermediate $p_{t}$ data (pions, kaons, protons, lambdas,
xis), and our combined approach (core + corona) describes well the
data (better than the differences between STAR and PHENIX results).
Lacking space, we just discuss a (typical) example: the nuclear modification
factor (AA/pp/number of collisions), for pions, kaons, protons, and
lambdas in central AuAu collisions at 200 GeV, see fig. \ref{cap:nmf}.%
For understanding these curves, we simply have a look at fig. \ref{cap:core-pp},
where we compare the core contributions from AuAu (divided by the
number of binary collisions) with pp. Since for very central collisions
the core dominates largely, the ratio of core to pp (the solid line
divided by the dotted one, in fig. \ref{cap:core-pp}) corresponds
to the nuclear modification factor. We discussed already earlier the
very different behavior of the core spectra (phase space decay) compared
to the pp spectra (string decay): pions are suppressed, whereas heavier
particles are favored. Or better to say it the other way round: the
production of baryons compared to mesons is much more suppressed in
string decays than in statistical hadronization.

\textbf{So what we observe here, is nothing but the very different
behavior of statistical hadronization (plus flow) on one hand, and
string fragmentation on the other hand. This completely statistical
behavior indicates that the low $p_{t}$ partons do not suffer energy
loss, they get completely absorbed in the core matter. }

\textbf{The $R_{cp}$ modification factors (central over peripheral)
are much less extreme than $R_{AA}$, since peripheral AuAu collisions
are a mixture of core and corona (the latter one being pp-like), so
a big part of the effect seen in $R_{AA}$ is simply washed out. }

\begin{figure}
\begin{center}\includegraphics[%
  scale=0.3,
  angle=270]{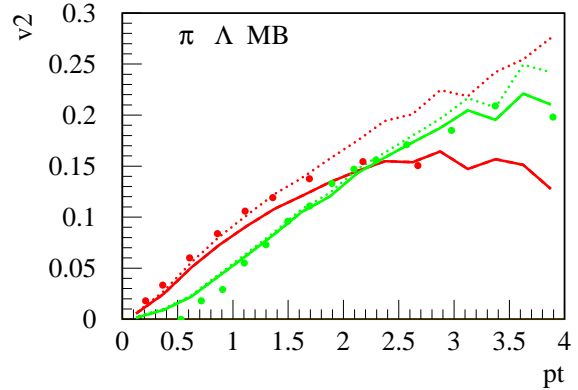}\end{center}
\vspace{-1.5cm}

\caption{Elliptical flow for pions (red) and lambdas (green), as a function
of transverse momentum. The points are data \cite{phenix-v2,star-v2-lda},
lines are calculations. The dotted lines represent only the core contribution,
the full lines are the complete contributions, core + corona.\label{cap:Ellip}}
\end{figure}

Let us finally discuss elliptical flow, shown in fig. \ref{cap:Ellip}.%
We understand the results in the following way: the pion curve seems
to saturate at high $p_{t}$, which is here simply due to the fact
that with increasing $p_{t}$ the continuously increasing core curve
is more and more {}``contaminated'' by corona contributions. For
the lambdas, the effect is much smaller, since the corona contributions
are smaller, as seen from fig. \ref{cap:core-corona}. Eventually,
the lambda curve will also saturate, but at larger $p_{t}$.

To summarize: we have discussed the influence of the corona contribution
(occurring in the periphery of nuclear collisions) in AuAu collisions
at RHIC. Our analysis is based on a model which works excellently
for pp and pA, together with a very simple parameterization of the
central (core) part. The fact that this simple treatment works, indicates
that the part we are really interested in, the core, shows a very
simple behavior. For example, contrary to the general believe, there
seems to be no centrality dependence of particle production, just
the volume changes.

We do not make any attempt here to explain these very interesting
data, the only purpose here is to separate the interesting part (core)
from the contamination (corona). We also did not make any efforts
to optimize the fits, actually most parameters are essentially first
guesses. To get more precision one need to enter into a more technical
discussion about for example the feed-down correction procedures in
the different experiments.


\begin{thebibliography}{1}
\bibitem{epos}K. Werner, Fuming Liu, Tanguy Pierog, J. Phys. G: Nucl. Part. Phys.
31 (2005) S985-S988.
\bibitem{phenix}S.S. Adler et al. (PHENIX Collaboration), submitted for publication
in PRC., nucl-ex/0307022. 
\bibitem{phenix-v2}PHENIX collaboration, Phys.Rev.Lett. 91 182301 (2003), nucl-ex/0305013 
\bibitem{star-v2-lda}STAR Collaboration, Phys. Rev. Lett. 92 (2004) 052302, nucl-ex/0306007
\bibitem{star-lda}M. Estienne (STAR Collaboration), J.Phys. G31 (2005) S873-S880\end{thebibliography}
\end{document}